# An investigation of the interrelationship between pressure and correlation in LaFeAsO under pressure

M. Hesani and A. Yazdani

*Department of Physics, University of Tarbiat Modares, Tehran, Iran*

## Abstract

Here, we investigated the interrelationship between pressure and correlation in the LaFeAsO compound by the density functional theory method combined with the dynamical mean field theory (DFT+DMFT) method. The spectral function and the occupation number (ON) of Fe-3d orbitals at different pressures were extracted from the calculations, and the importance of the role of correlation in superconductivity in iron-based compounds was indicated. The measured ON of Fe-3d orbitals demonstrated charge transfer between them and also the crucial role of the Fe-3dxy (xy) orbital in superconductivity in these materials. A switching behavior between high and low TC through decreasing the correlation was observed in our calculations, originating from the orbital switching. This shows the significant role of the orbital degrees of freedom in iron-based superconductors.

## 1. Introduction

Since the discovery of superconductivity in the iron-based compound of LaO1-xFxFeAs with a transition temperature (Tc) of 26 K [1], this branch of research becomes one of the most exciting fields in physics. Several parent compounds of these materials, such as LaFeAsO, BaFe2As2, LiFeAs, and FeSe etc., have been found and explored in these years [2–5]. Chemical substituting (i.e., doping with F [1] or H [6]) and applying a high pressure [7] on the parent compound (i.e. LaFeAsO) lead to superconductivity (SC) in this class of materials. The pressure significantly alters the magnetic, structural, electronic, and superconductivity properties of these systems [8–17]. It seems that studying of pressure in iron-based compounds is very important for finding the cause of superconductivity in them.

In these materials, $T_c$ is very sensitive to the application of external pressures on them. Applying an external pressure on $LaFeAsO_{1-x}F_x$ [18] and $LaFeAsO_{1-x}H_x$ [19] increases $T_c$ to 43 and 52 K, respectively. The increase is caused by the enhancement of the charge transfer between the insulating and conducting layers [13,18,20]. Furthermore in $LaFeAsO_{1-x}H_x$, the $T_c$ valley at x=0.21 in the double-dome structure disappears at the ambient pressure, resulting in a single-dome structure with a higher $T_c$ [21]. The application of a high pressure on the parent compound of LaFeAsO suppresses the magnetic ordering at 20 GPa and changes the crystal structure at 30 GPa [8]. A quantum phase transition from the antiferromagnetic phase (AF) to the SC one occurs in the heavily hydrogen-doped $LaFeAsO_{1-x}H_x$ compound under the 3 GPa pressure [17]. A density functional theory (DFT) study on the undoped LaFeAsO compound shows that the Fe moment drops by a factor of ~3 within the pressure range of ±5 GPa [22]. It also predicts that the magnetic order disappears under 20 GPa [22] or 29.5 GPa [23] pressures.

The exploration of the electronic structure of these materials under pressure may be very helpful in understanding high Tc iron-based superconductors.

The dome-shaped pressure dependence of TC in the LaFeAsO compound is similar to the chemical doping dependence; TC increases under pressure to reach its maximum (21 K) at 12 GPa and then decreases [7,8,24]. The anionic or pnictogen height (hAs), which is the distance of the As atom from the Fe layer, decreases as the pressure increases [8]. It has been reported that the optimum TC is found when the FeAs4 tetrahedron is close to the regular shape and the anionic height is close to 1.38 A° for iron pnictides [25]. Kumar et al. [8] found the optimum TC at 1.34 Å in LaFeAsO under pressure. Furthermore, it has been reported that an iso-structural superconductivity transition with H doping in LaFeAsO with a sizeable change in the pnictogen height increases TC [6]. This shows the sensitivity of both chemical doping and the applied pressure to the anionic height, which has a close relationship with correlation in these materials [9,26–28]. Kuroki et al [29] showed that the pnictogen height plays an important role as a switch between high TC nodeless and low TC nodal pairing. The s-wave pairing dominates for high pnictogen heights and the reduction of hAs results in the d-wave pairing. This suggests that the correlation between the effect of pressure on superconductivity is very important and crucial for tailoring new high TC compounds based on Fe pnictides.

The density functional theory method combined with the dynamical mean field theory method (DFT+DMFT) is very successful in predicting the magnetic, structural, electronic, and superconductivity properties of these systems [30–33]. Its emphasis is on the crucial role of correlation in these materials.

Here, we investigated the interrelationship between pressure and correlation in LaFeAsO compounds by the density functional theory method combined with the dynamical mean field theory method (DFT+DMFT). The spectral function and the occupation number of Fe-3d orbitals at different pressures were extracted from calculations, and the importance of the role of the Fe-3dxy (xy) orbital in superconductivity in this materials under pressure was indicated.

## 2. Method and materials

We used the charge self-consistent combination of density-functional theory and dynamical mean field theory (DFT+DMFT)[34], which is implemented in two full-potential methods, the augmented plane-wave and the linear muffin-tin orbital methods (WIEN2K code [35]). DFT calculations were done in the WIEN2K package [27] in its generalized gradient approximation [Perdew-Burke-Ernzerhof (PBE)-GGA] [36]. The self-energy is added to the DFT output, and the DMFT was run. The computations are converged with respect to the charge density, total energy, and self-energy. The continuous-time quantum Monte Carlo (CTQMC) method [37,38] in its fully rotationally invariant form was used to solve the quantum impurity problem. All 3d orbitals, d3z2-r2 (z2), dx2-y2 (x2y2), dxz (xz), dyz (yz), and dxy (xy), were considered as correlated orbitals for the Fe site. The self-energy was analytically continued from the imaginary axis to the real axis using the maximum entropy method.

The experimentally determined lattice structures, including the internal positions of the atoms for the LaFeAsO compound at different pressures and a low temperature, were used for these calculations. The experimental (exp) lattice structures were taken from the work of Kumar et al.

[8] (their supplementary information), as presented in **Error! Reference source not found.**. In all DFT + DMFT calculations, the same Coulomb interaction U = 5.0 eV and Hund's coupling JH = 0.80 eV were used [30]. A fine k-point mesh of 16 × 16 × 7 and totally 3×10$^9$ Monte Carlo steps for each iteration were used for the non-magnetic phase of LaFeAsO compound at the different pressures at 150 K. Structural positions were relaxed by DFT and DFT + DMFT methods and the results are reported in **Error! Reference source not found.**. The DFT relaxations were done for the non-magnetic (NM) and the stripe-type antiferromagnetic (AF) states. The DFT+DMFT relaxations were done by the method reported by Kristjan Haule and Gheorghe L. Pascut for NM states [33].

**Table 1.** Structural parameters of LaFeAsO at various pressures and a comparison between experimentally reported structures and our optimized structures (obtained using DFT and DFT+DMFT).

| Parameter | 0 GPa | 0.5 GPa | 3.3 GPa | 6.3 GPa | 9.2 GPa | 13.4 GPa | 14.5 GPa | 20 GPa | 34 GPa |
|---|---|---|---|---|---|---|---|---|---|
| a (Å) | 4.0353 | 5.7221 | 5.6982 | 5.6719 | 5.6547 | 5.6282 | 5.6238 | 5.5848 | 5.4314 |
| b (Å) | 4.0353 | 5.7170 | 5.6762 | 5.6499 | 5.6114 | 5.5658 | 5.5622 | 5.5237 | 5.4314 |
| c (Å) | 8.7409 | 8.7419 | 8.6327 | 8.5661 | 8.4512 | 8.3606 | 8.3452 | 8.2057 | 7.7860 |
| As z (exp) | 0.6512 | 0.6651 | 0.6620 | 0.6563 | 0.6554 | 0.6545 | 0.6549 | 0.6566 | 0.6567 |
| (DFT+DMFT relax) | 0.6564 | 0.6582 | 0.6588 | 0.6580 | 0.6598 | 0.6609 | 0.6611 | 0.6623 | 0.6665 |
| (DFT relax, NM) | 0.6381 | 0.6384 | 0.6402 | 0.6410 | 0.6428 | 0.6445 | 0.6448 | 0.6468 | 0.6533 |
| (DFT relax, AF) | 0.6473 | 0.6476 | 0.6481 | 0.6483 | 0.6489 | 0.6495 | 0.6497 | 0.6507 | 0.6548 |
| La z (experimental) | 0.14154 | 0.1409 | 0.1475 | 0.1497 | 0.1513 | 0.1519 | 0.1524 | 0.1500 | - |
| (DFT+DMFT relax) | 0.1412 | 0.1404 | 0.1422 | 0.1435 | 0.1450 | 0.1466 | 0.1468 | 0.1488 | 0.1546 |
| (DFT relax, NM) | 0.1441 | 0.1433 | 0.1451 | 0.1462 | 0.1478 | 0.1493 | 0.1495 | 0.1514 | 0.1573 |
| (DFT relax, AF) | 0.1427 | 0.1424 | 0.1442 | 0.1455 | 0.1472 | 0.1488 | 0.1490 | 0.1510 | 0.1570 |
| h(As) (Å) (exp) | 1.32 | 1.44 | 1.40 | 1.34 | 1.31 | 1.29 | 1.29 | 1.285 | 1.22 |
| (DFT+DMFT relax) | 1.32 | 1.38 | 1.37 | 1.35 | 1.35 | 1.34 | 1.34 | 1.33 | 1.29 |
| (DFT relax, NM) | 1.21 | 1.21 | 1.21 | 1.21 | 1.21 | 1.21 | 1.21 | 1.20 | 1.19 |
| (DFT relax, AF) | 1.29 | 1.29 | 1.28 | 1.27 | 1.26 | 1.25 | 1.25 | 1.24 | 1.21 |
| d(FeAs-LaO) (Å) (exp) | 1.8116 | 1.6959 | 1.6445 | 1.6618 | 1.6336 | 1.6186 | 1.6081 | 1.5870 | - |
| (DFT+DMFT relax) | 1.7693 | 1.7602 | 1.7175 | 1.700 | 1.6494 | 1.6100 | 1.6032 | 1.5501 | 1.3920 |
| (DFT relax, NM) | 1.9039 | 1.9079 | 1.8535 | 1.8225 | 1.7695 | 1.7235 | 1.7166 | 1.6551 | 1.4718 |
| (DFT relax, AF) | 1.8356 | 1.8360 | 1.7931 | 1.7664 | 1.7232 | 1.7044 | 1.7014 | 1.6268 | 1.5530 |

## 3. Results and discussion

The optimized structural positions calculated by DFT and DFT+DMFT methods are reported in Table 1. The DFT results for the NM states for the structural parameters are much different from the experimental measurements. The anion height (h(As) is approximately constant at different pressures, which is not consistent with measurements. These materials are very sensitive to the anion height, which shows that we have to apply methods beyond DFT to these materials. The distance between FeAs and LaO planes is overestimated by the DFT method.

Decreasing the parameter facilitates the charge transfer between layers and is a very important parameter in superconductivity. The relaxed structures by the DFT method for the AF state are improved significantly, but a method beyond DFT such as DFT+DMFT could make the results much better. Our DFT+DMFT results have a better agreement with experimental measurements. The anion height shows the processes reported by experimental measurements and the distance between FeAs and LaO planes has a good agreement with experimental measurements. Our results show the strength of the DFT+DMFT method in predicting the structural parameters in these materials.

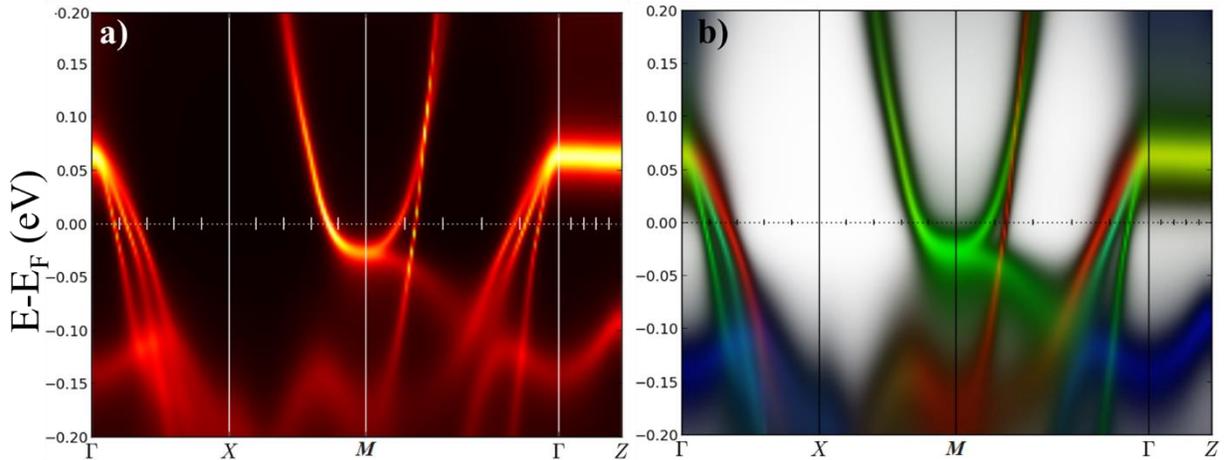

**Figure 1.** (Color online). a) The computed DFT+DMFT spectral function and b) its corresponding orbital resolved spectral function of the LaFeAsO compound at ambient pressures. z2 and x2−y2 are in blue, dxz and dyz are in green, and dxy is in red.

The computed DFT+DMFT spectral function of the LaFeAsO compound at the ambient pressure and its corresponding orbital resolved spectral function are plotted in Figure 1. For this calculation, we used the structural parameter reported by Kamihara et al. [1]. The spectral function shows three hole pockets at the center of the Brillouin zone and two electron pockets at the zone corner, which is confirmed by previous studies [28,39]. At the center, the outer band has mostly the xy character. The middle band is a combination of xz/yz and x2y2 characters, and the inner band is mostly composed of the xz/yz character, as it is illustrated in Figure 1-b. The mass enhancements were calculated as 2.35, 2.66, and 2.75 for x2y2/z2, xz/yz, and xy orbital, respectively, which have a good agreement with Yin et al. work [30].

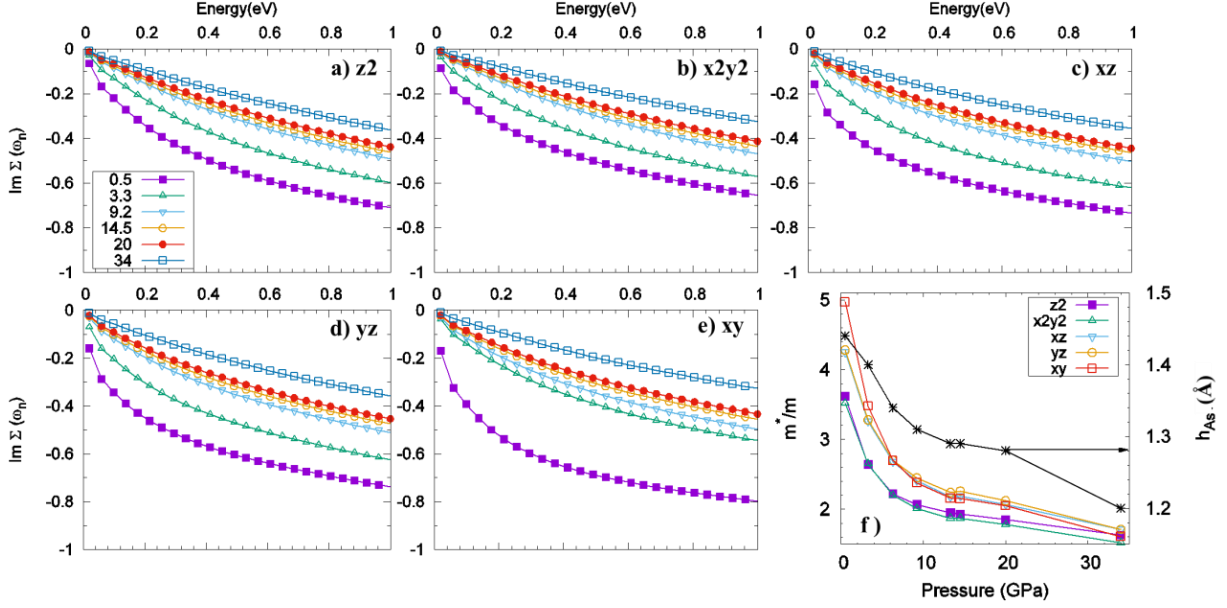

**Figure 2.** (Color online) The pressure-dependent imaginary part of self-energy (ImΣ(iωn)) of Fe-3d orbitals at a low energy, (a) dz2, (b) dx2−y2, (c) dxz, (d) dyz, (e) dxy and mass enhancement (f) m*/m at 150 K. In figure (f), the mass enhancement in different Fe-3d orbitals (left) and the anionic height ($h_{As}$) (right) showed a similar trend across eight different pressures. The black vector at (f) points to the corresponding axis.

The pressure-dependent imaginary part of the self-energy ($\mathrm{Im}\Sigma(i\omega_n)$) of Fe-3d orbitals is shown in Figure 2, which shows that ImΣ is very orbital selective. As the pressure increases, the ImΣ behavior at low energies becomes more linear, with a slope directly related to the quasiparticle mass enhancement, as expected from the Fermi liquid theory. Increasing pressure enhances the coherence scale in these materials. In the Fermi liquid theory, the inverse quasiparticle (QP) lifetime is defined as the scattering rate $\Gamma = -Z\, \mathrm{Im}\Sigma(i\omega)$, where $Z = (m^*/m)^{-1} = (1 - \frac{\partial \Sigma(i\omega)}{i\omega})^{-1}|_{i\omega \to i0^+}$ can be interpreted as the quasiparticle weight and $\mathrm{Im}\Sigma(i0^+)$ is the imaginary part of the self-energy at the zero frequency. At the zero temperature, when $\mathrm{Im}\Sigma(i0^+) \to 0$, the system is in the coherent phase with an infinite QP lifetime. At a constant temperature, ImΣ(i0+) and consequently the QP lifetime are both finite. Therefore, in order to identify the coherent QPs, the temperature has to be lower than the QP width, i.e. $T < -Z\, \mathrm{Im}\Sigma(i0^+)$. At P=3.3 GPa, we have $-\mathrm{Im}\Sigma(i0^+)_{xy} \simeq 0.07$ eV and Zxy=0.29 that give $\Gamma_{xy}$ corresponding to 236 K, which is higher than 150 K and, hence, no coherent quasi-particle can be expected. For xz and yz orbitals, the QP width corresponds to a temperature more than 230 K, which makes them as the former. For x2y2 and z2 orbitals, $-Z\, \mathrm{Im}\Sigma(i\omega) \simeq 160$ k that is as order of our calculation, so nearly coherent phase is expected for these orbitals against the others. However, for P higher than 3.3 GPa, the system is expected to be at the coherent metal phase. For P lower than 3.3 GPa, a temperature lower than 150 K is needed for the coherent phase. Therefore, the pressure induces a transition from the incoherent state to the coherent state in the LaFeAsO compound.

The variation of orbital-dependent effective masses, as shown in Figure 2 (f) follows the variation of anionic height at different pressure, which confirms the sensitivity of correlation in these materials to the anionic height. The increase of the anionic height higher than 1.40 Å influences severely on the correlation and enhances the effective mass to 5 for the xy orbital. The highest variation in the mass enhancement is for the xy orbital. It changes from 5 at 0.5 GPa to 1.6 at 34 GPa. The decrease of correlation in these materials helps in increasing the superconductivity temperature (TC) to reach its maximum at 12 GPa [8], and then decreases as P increases more. In our calculation, the optimum P is next to 13.3 GPa. The mass enhancement variation in the range of 13.3 to 20 GPa is very smoothly. This suggests that the other parameters play a role in decreasing TC in this region.

**Table 2.** The occupation number of Fe-3d orbitals at different pressures.

| P (GPa)\ON | n-impurity | z2 | x2y2 | xz | yz | xy |
|---|---|---|---|---|---|---|
| 0.5 | 6.25 | 1.33 | 1.24 | 1.22 | 1.22 | 1.24 |
| 3.3 | 6.273 | 1.32 | 1.20 | 1.23 | 1.23 | 1.30 |
| 6.3 | 6.284 | 1.32 | 1.17 | 1.24 | 1.24 | 1.32 |
| 9.2 | 6.281 | 1.32 | 1.15 | 1.24 | 1.24 | 1.33 |
| 13.3 | 6.299 | 1.33 | 1.14 | 1.25 | 1.24 | 1.34 |
| 14.5 | 6.298 | 1.33 | 1.14 | 1.25 | 1.24 | 1.34 |
| 20 | 6.297 | 1.33 | 1.14 | 1.25 | 1.24 | 1.34 |
| 34 | 6.31 | 1.33 | 1.10 | 1.27 | 1.27 | 1.35 |

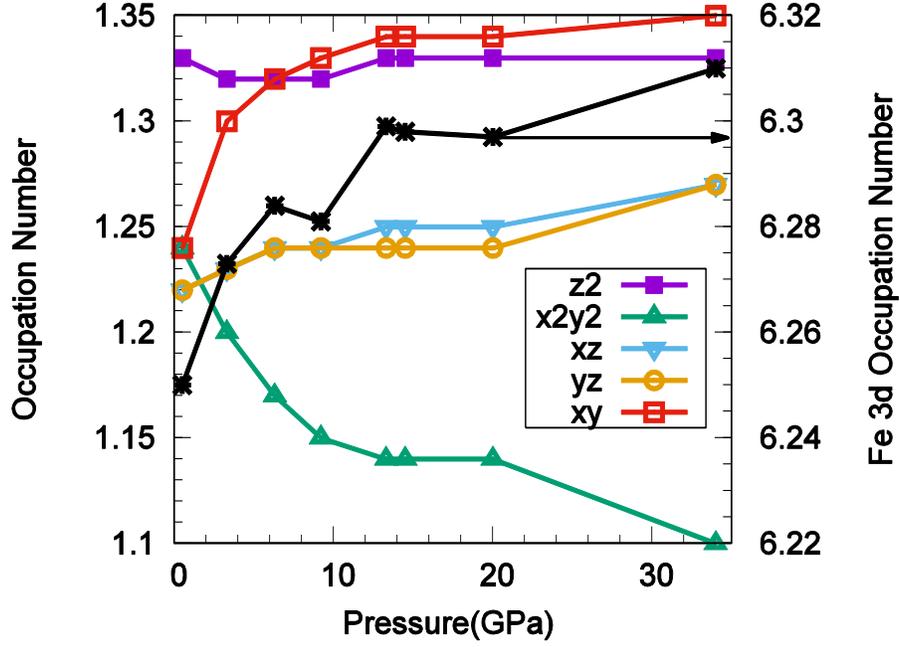

**Figure 3.** (Color online) The occupation number of Fe-3d orbitals at different pressures (left). The Fe-3d occupation number (right). The black vector points to the corresponding axis.

To monitor the variation of Fe-3d orbitals, we computed the occupation number (ON) for Fe-3d orbitals at different pressures. Table 2 and Figure 3 present the variation of the ON of these orbitals at different pressures. The enhancement of the carrier concentration in 3d orbitals by the applied pressure may be caused by the increase of the charge transfer between the inter- and intra-planes through the lattice compressions. As determined from our calculations, the highest increase and decrease of the ON is for the xy and the x2y2 orbitals, respectively. The variation of the ON for three orbitals of xz, yz, and z2 are insignificant for pressures in the range of 0.5 to 14.5 GPa. It seems that by applying a pressure, we have charge transfer between Fe-3d orbitals. The increase P to 14.5 GPa increases (decreases) the ON about 0.1 in the xy (x2y2) orbital. This phenomenon guides the system towards the highest TC. This suggests that the charge transfer between orbitals plays an important role in TC in these materials under pressure. The ON variation in the xy orbital is insignificant for P higher than 13.3 GPa. At pressures higher than 14.5 GPa, the xz/yz orbital plays the role of the xy orbital and ON are increased in them. The fluctuation of the ON at P higher than 14.5 GPa is not in favor of increasing TC. The issue would be discussed more in the following. A switching behavior from the xy orbital as the acceptor to the xz/yz orbital as the acceptor inverses the slope of the dome-shaped pressure dependence of TC. This explains the microscopic origin of the dome-shaped pressure dependence of TC and pointed the importance of correlation in superconductivity in these materials.

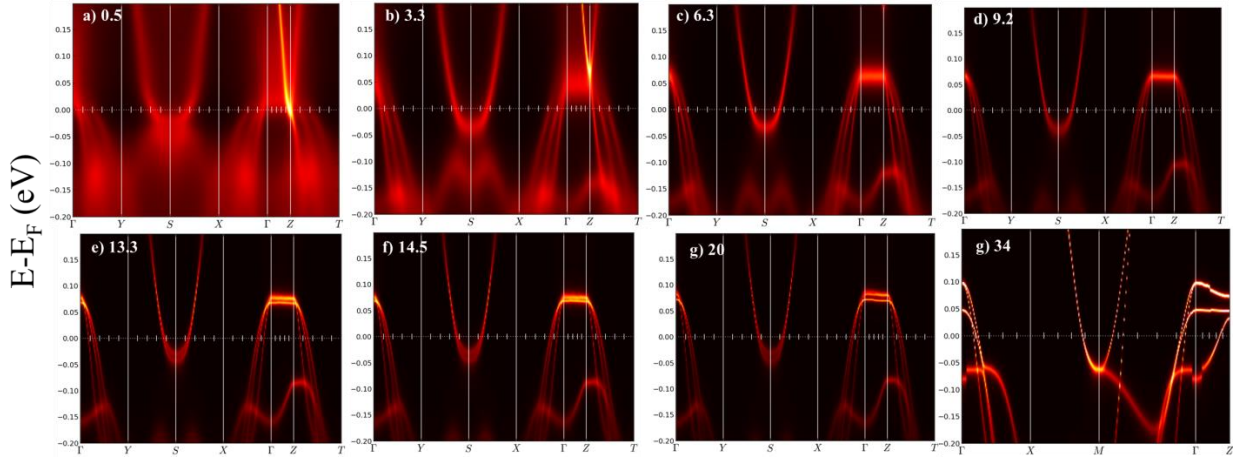

**Figure 4.** (Color online). The computed DFT+DMFT spectral function of the LaFeAsO compound at different pressures, a) 0.5, b) 3.3, c) 6.3, d) 9.2, e) 13.3, f) 14.5, g) 20 and h) 34 GPa.

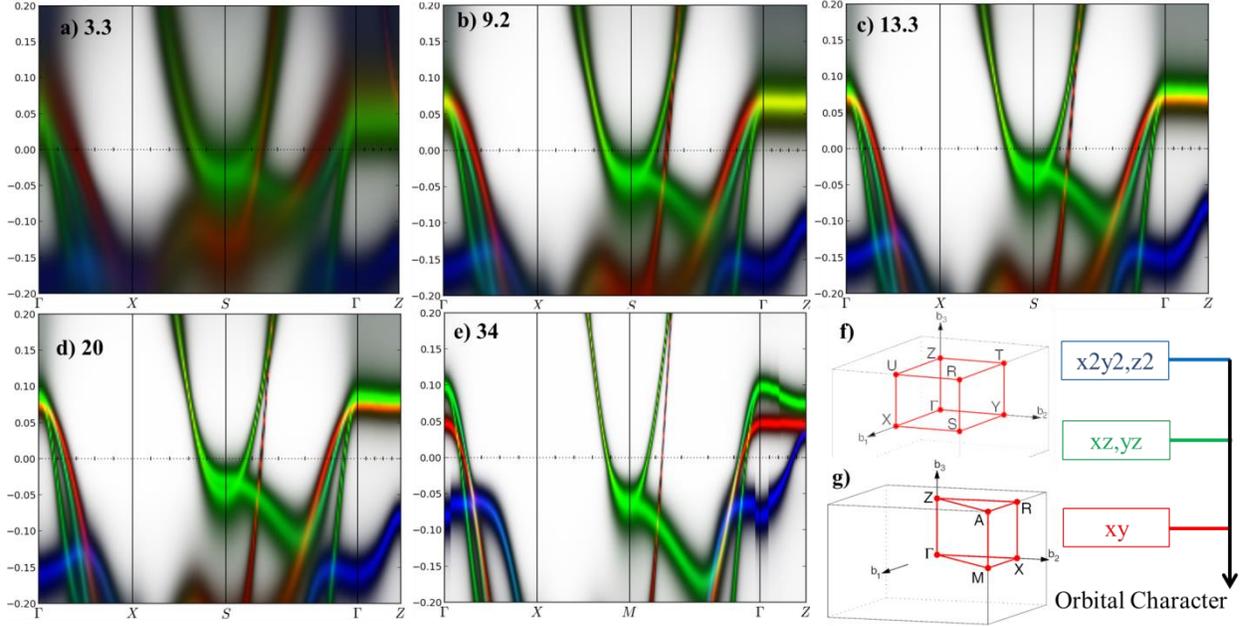

**Figure 5.** (Color online). The corresponding orbital resolved DFT+DMFT spectral function of the LaFeAsO compound at different pressures, a) 3.3, b) 9.2, c) 13.3, d) 20, and e) 34 GPa. f) and g) Brillouin zones of simple orthorhombic and simple tetragonal lattice structures, respectively. z2 and x2−y2 are in blue, dxz and dyz are in green, and dxy is red in the figure.

The computed DFT+DMFT spectral function of the LaFeAsO compound at different pressures and its corresponding orbital resolved spectral function are plotted in Figure 4 and Figure 5, respectively. The spectral function at 0.5 and 3.3 GPa are not very clear, and hence it confirms our discussion that at these pressures, the spectral function is not at its coherent state. As the pressure increases, the spectral function goes to its coherent state, and it becomes more clear. At a low pressure, the results show three hole pockets at the center of the Brillouin zone and two electron pockets at the zone corner, which is in agreement with the spectral function of the compound at the ambient pressure, as discussed before. At the center, the outer band has mostly the xy character. The middle band is a combination of xz/yz and x2y2 characters, and the inner band is mostly composed of the xz/yz character, as it is illustrated in Figure 5. As the pressure increases, the outer pocket becomes smaller, but the middle and inner pockets remain nearly unchanged. It means that the xy orbital gets some weight and its hole state decreases. In other word, the ON of the xy orbital gains some weight, which is confirmed by our previous discussions. At pressures higher than 14.5 GPa, the outer pocket variation is insignificant, as opposed to the middle pocket that starts to grow. This is mostly because of the x2y2 orbital that loses some weight, as shown before. At 34 GPa, the middle pocket nearly becomes larger than the outer pocket at some point, as shown in Figure 6. At pressures higher than 14.5 GPa, the degeneracy in the system increases, as shown in Figure 5, which is not in favor of increasing TC. Some authors pointed the same results in their research[40], which is caused by doping more F in LaFeAsO1-xFx than its optimum.

At the zone corner, two electron pockets create the Fermi surface (FS). The outer band has mostly the xy character and the inner band has the xz/yz character. As the pressure increases, the outer pocket gains some weight and grows larger. It helps in better nesting in the FS. Our calculation shows that at a high pressure, the size of the hole (electron) pocket decreases (increases) and this phenomenon makes a better nesting in the FS until the system reaches its highest TC. The main role in this phenomenon is played by the xy orbital. This suggests the importance of the fluctuations mediated by the disconnected nesting in the high TC superconductors.

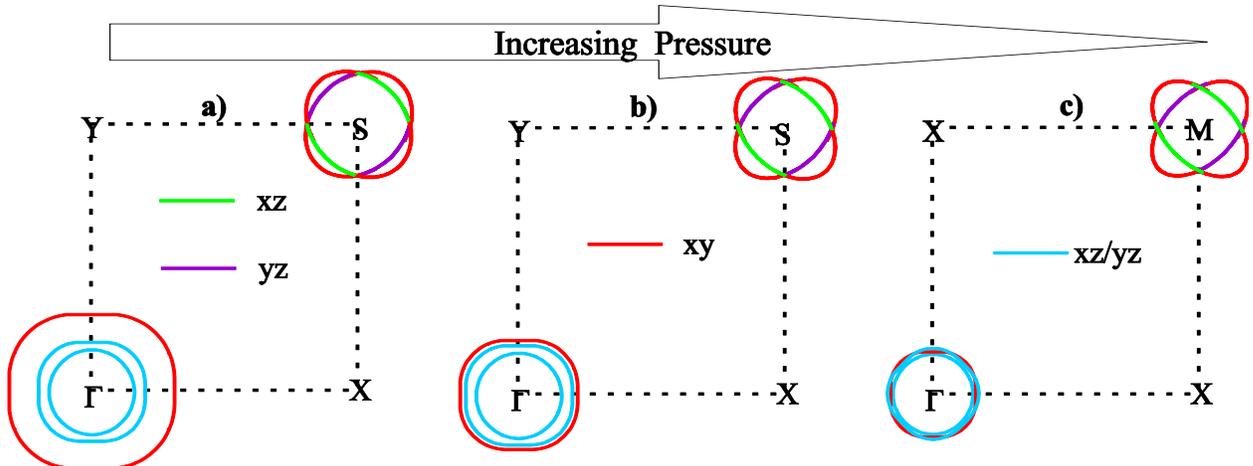

**Figure 6.** (Color online) The schematic of the Fermi surface variation by increasing the applied pressure.

A schematic of the Fermi surface variation under a high pressure is illustrated in Figure 6. The schematic shows the process of FS variation under pressure, deriving from Figure 5. Colors show the orbitals that mostly characterized the pockets. Part (a) is extracted from the spectral function at 3.3 GPa (Figure 5-a). The differences between parts (a) and (b) of Figure 6 are mostly induced by variation of the xy orbital. It is in favor of the increase of TC. Figure 6-b is extracted from Figure 5-c, which is at the pressure of 13.3 GPa. This pressure is close to the optimum pressure. In Figure 6-c, the schematic of the FS of the compound at the pressure of 34 GPa is represented. The x2y2 orbital is the main factor of the FS variation over the pressure range of 13.3 to 34 GPa. It is not in favor of the increase of TC, and gets the system away from its optimum.

Our calculations demonstrated charge transfer between Fe-3d orbitals in the system, and the origins of the phenomenon are different in either sides of the optimum pressure. Before the optimum pressure, the fluctuations originate from the xy orbital and after that, they originate from the xz/yz orbital. Furthermore, the charge transfer varies the fermiology of the system. This phenomenon may result from the disconnected nesting at the FS. Kuroki et al. [41] proposed multiple spin fluctuations arising from the disconnected FS as the origin of superconductivity in LaFeAsO compounds. Our results show that if it is the cause of superconductivity, the contribution of the orbital degrees of freedom must not be ignored.

Kuroki et al. [29] showed that the pnictogen height plays an important role as a switch between the high TC nodeless pairing and the low TC nodal pairing. The s-wave pairing dominates for high pnictogen heights, and reduction of hAs results in the d-wave pairing. This result is derived from the same assumption as above. A switching behavior between high and low TC through decreasing the pnictogen was observed in our calculations, originating from the orbital switching, as discussed before. This finding may confirm the work of Kuroki et al.; however, here, the orbital degrees of freedom play an important role in this theory.

## 4. Conclusions

In summary, we calculated the electronic structure of the LaFeAsO compound under pressure by DFT+DMFT methods. Our results showed a transition from incoherent to coherent states when applying pressure. The decrease of correlation when the pressure increases was obvious in the mass enhancement graph. The measured ON of Fe-3d orbitals demonstrated the fluctuation of electrons between them. At first, mostly the xy orbital plays the role of the acceptor to reach its optimum. Then, this role mostly switches to the xz/yz orbital in the area where TC of the system decreases. At all pressures, nearly the role of the donor is played by the x2y2 orbital. This suggest the crucial role of correlation, which is the cause of the switching between orbitals. Our findings may provide a new route towards a better understanding of high TC superconductors. Increasing the pressure alters the fermiology of the system. It helps in achieving a better nesting at the optimum pressure and then increasing the degeneracy of the system, which occurs mostly at the area where system gets way from its optimum. It seems that the degeneracy is not in favor of increasing TC in these materials. A switching behavior between high and low TC through decreasing the pnictogen was observed in our calculations, originating from the orbital switching. This shows the significant role of the orbital degrees of freedom in iron-based superconductors.

### Acknowledgement

We are very grateful to the professor Kristjan Haule for providing the DFT+DMFT code and also his tutorials and responses to our questions and comments. We also thank Dr. Gheorghe L. Pascut for his tutorials and comments.